\documentclass[twocolumn]{aastex631}

\usepackage{natbib}
\usepackage{multirow}
\usepackage{booktabs}
\usepackage{epsfig}
\usepackage{ulem}
\usepackage{rotating}
\usepackage{graphicx}
\usepackage{url}
\usepackage{xcolor}
\usepackage{amsmath}
\usepackage{color}
\usepackage{savesym}
\savesymbol{tablenum}
\usepackage{siunitx}
\usepackage[normalem]{ulem}
\usepackage{appendix}
\usepackage{enumerate}
\usepackage{gensymb}
\usepackage{longtable}
\usepackage{afterpage}
\usepackage{threeparttable}
\usepackage{tabularx}
\usepackage[bottom]{footmisc}

\newcommand\hGpc{$h^{-1}$~Gpc}
\newcommand\hMpc{$h^{-1}$~Mpc}
\DeclareSIUnit\parsec{pc}
\DeclareSIUnit\h{\textit{h}}

\begin{document}
    \title{Improving the Determination of Supernova Cosmological Redshifts by Using Galaxy Groups}
    \author[0000-0001-8596-4746]{Erik R.~Peterson}
    \affiliation{Department of Physics, Duke University, Durham, NC 27708, USA}
    \author[0000-0002-7234-844X]{Bastien Carreres}
    \affiliation{Department of Physics, Duke University, Durham, NC 27708, USA}
    \author[0000-0003-4074-5659]{Anthony Carr}
    \affiliation{School of Mathematics and Physics, University of Queensland, Brisbane, QLD 4072, Australia}
    \affiliation{Korea Astronomy and Space Science Institute, Yuseong-gu, Daedeok-daero 776, Daejeon 34055, Republic of Korea}
    \author[0000-0002-4934-5849]{Daniel Scolnic}
    \affiliation{Department of Physics, Duke University, Durham, NC 27708, USA}
    \author{Ava Bailey}
    \affiliation{Department of Physics, Duke University, Durham, NC 27708, USA}
    \author[0000-0002-4213-8783]{Tamara M.~Davis}
    \affiliation{School of Mathematics and Physics, University of Queensland, Brisbane, QLD 4072, Australia}
    \author[0000-0001-5201-8374]{Dillon Brout}
    \affiliation{Department of Astronomy and Physics, Boston University, Boston, MA 02140, USA}
    \author[0000-0002-1081-9410]{Cullan Howlett}
    \affiliation{School of Mathematics and Physics, University of Queensland, Brisbane, QLD 4072, Australia}
    \affiliation{OzGrav:~The ARC Centre of Excellence for Gravitational Wave Discovery, Hawthorn, VIC 3122, Australia}
    \author[0000-0002-6230-0151]{David O.~Jones}
    \affiliation{Institute for Astronomy, University of Hawai`i, 640 N.~A'ohoku Pl., Hilo, HI 96720, USA}
    \author[0000-0002-6124-1196]{Adam G.~Riess}
    \affiliation{Space Telescope Science Institute, Baltimore, MD 21218, USA}
    \affiliation{Department of Physics and Astronomy, Johns Hopkins University, Baltimore, MD 21218, USA}
    \author[0000-0002-1809-6325]{Khaled Said}
    \affiliation{School of Mathematics and Physics, University of Queensland, Brisbane, QLD 4072, Australia}
    \author[0000-0001-5756-3259]{Georgie Taylor}
    \affiliation{Research School of Astronomy and Astrophysics, The Australian National University, ACT 2601, Australia}

\begin{abstract}

At the low-redshift end ($z<0.05$) of the Hubble diagram with Type Ia Supernovae (SNe Ia), the contribution to Hubble residual scatter from peculiar velocities is of similar size to that due to the limitations of the standardization of the SN Ia light curves. 
A way to improve the redshift measurement of the SN host galaxy is to utilize the average redshift of the galaxy group, effectively averaging over small-scale/intracluster peculiar velocities.
One limiting factor is the fraction of SN host galaxies in galaxy groups, previously found to be 30\% using (relatively incomplete) magnitude-limited galaxy catalogs. 
Here, we do the first analysis of N-body simulations to predict this fraction, finding $\sim$73\% should have associated groups and group averaging should improve redshift precision by $\sim$135 km~s$^{-1}$ ($\sim$0.04 mag at $z=0.025$). 
Furthermore, using spectroscopic data from the Anglo-Australian Telescope, we present results from the first pilot program to evaluate whether or not 23 previously unassociated SN Ia hosts belong in groups.
We find that 91\% of these candidates can be associated with groups, consistent with predictions from simulations given the sample size. 
Combining with previously assigned SN host galaxies in Pantheon+, we demonstrate improvement in Hubble residual scatter equivalent to 145 km~s$^{-1}$, also consistent with simulations. 
For new and upcoming low-$z$ samples from, for example, ZTF and LSST, a separate follow-up program identifying galaxy groups of SN hosts is a highly cost-effective way to enhance their constraining power.

\end{abstract}

\section{Introduction}

Type Ia Supernovae (SNe Ia) play a critical role in constraining key cosmological parameters such as the Hubble constant $H_0$, the dark energy equation-of-state parameter $w$, and the growth-of-structure parameter $f\sigma_8$ among others \citep[e.g.,][]{Brout22,Carreres23,DES24,DESI24}.
The precision of measurements on these cosmological parameters depends heavily on how well we are able to leverage SNe Ia as distance indicators.
Specifically, the accuracy of both SN distances and cosmological redshifts are important for obtaining both accurate and precise cosmological parameters.

Peculiar velocities (PVs), defined as motions of galaxies with respect to the cosmological rest frame, 
can be a statistical and systematic uncertainty in analyses of the Hubble diagram \citep[e.g.,][]{Peterson22,Carreres24}.
Combined, these contribute around $\sim$250--300 km s$^{-1}$ of uncertainty from motion on different physical scales.
In terms of systematics for constraining cosmological parameters, larger-scale (10s of \hMpc{}) correlations are the most important.
For example, \citet{Peterson22} provide evidence that correcting for larger-scale PVs improves scatter on the Hubble diagram and can shift $H_0$ by up to $\sim$0.4 km s$^{-1}$ Mpc$^{-1}$.
With Zwicky Transient Facility (ZTF) simulations, \citet{Carreres24} find that not accounting for the full covariance between SNe due to bulk-flow PVs can shift the best-fit $H_0$ of a ZTF-like sample by up to 1~km~s$^{-1}$ Mpc$^{-1}$.
For statistical uncertainties, smaller-scale ($<$10 \hMpc{}) PVs become more important; \citet{Peterson22} find that accounting for small-scale motions by associating redshifts with galaxy groups results in large improvements in $\chi^2$ values (up to 37\%) of Hubble residuals.

While \citet{Peterson22} show that group assignments improve Hubble residual scatter, they find that only 30\% of their sample of low-$z$ SN host galaxies from the Pantheon+ analysis \citep{Brout22,Scolnic22} can be assigned to galaxy groups. 
However, the group catalogs they use are magnitude limited, which can be illustrated by the fraction in groups dropping from 49.6\% for 0.01$<$\textit{z}$<$0.02 to 9.6\% for 0.04$<$\textit{z}$<$0.05. 
Each galaxy group catalog defines galaxy groups slightly differently, but all methods are based on associating velocities and distances among galaxies.
\citet{Crook07} analyze an early release of the 2MRS \citep{Huchra05a,Huchra05b} catalog, and of the $>$20,000 galaxies in their sample, they deem that 72.8\% of them are in groups of size two or larger.
For the group catalog from \citet{Tully15} (hereafter T15), out of the 24,044 galaxies in the analysis sample from a later release of 2MRS \citep{Huchra12}, 
13,900 of those galaxies are in groups with at least two members, a rate of 57.8\%. 
Although SNe Ia tend to be hosted by galaxies that are on the more massive and more star-forming end of the full distribution of galaxies in our universe \citep{Smith12,Wiseman21}, this difference in galaxy mass distribution may not be enough to explain this discrepancy between the rate of galaxies in groups found by \citet{Peterson22} and these group catalogs.

The two main questions we aim to answer in this paper are (i) what the real fraction of SN host galaxies in groups is, and (ii) how much and how often group association improves Hubble residuals.
We take a two-pronged approach to these questions by using both simulations to make predictions and then new data to investigate them.
In terms of simulations, an analysis using simulations to predict either the rate of SN host galaxies in groups or the frequency at which galaxy groups improve Hubble residuals has yet to be done.
With the recent N-body simulations of our universe from Uchuu \citep{Ishiyama2021,Aung23}, 
a friends-of-friends \citep[FoF;][]{HuchraGeller82} group finder, 
and an accurate SN host map \citep[i.e.,][]{Sullivan06,Wiseman21},
we are able to execute an accurate galaxy group analysis of SN host galaxies and make predictions with simulations.
In terms of data, the observing requirements to obtain galaxy-group data for these host galaxies are relatively low. 
The galaxies targeted with such a program should be at low redshift ($z\lesssim0.06$; since PVs have the largest impact at low redshift) and thus bright.
Additionally, newly defined galaxy groups can be combined with Pantheon+ data to supplement sample statistics for analysis.

In this paper, (i) we use a simulated universe from Uchuu to make predictions on the rate of SN hosts in groups and the advantage of using galaxy-group information in SN analyses, and (ii) we carry out our own spectroscopic observations and analysis of a sample of SN host galaxies that have not been assigned to galaxy groups and combine with data from Pantheon+ to test these same predictions.
In Section~\ref{sec:data} we describe the data acquisition and data sample, and in Section~\ref{sec:galaxygroups} the definitions of groups from both simulations and data are given.
Results from the galaxy groups themselves are provided in Section~\ref{sec:GGresults}, while results on Hubble residuals are in Section~\ref{sec:HRresults}.
Finally, in Sections~\ref{sec:discussions} and~\ref{sec:conclusions} we present our discussions and conclusions.

\begin{table*}[!hbt]
\caption{SN Ia host galaxy sample observed with the AAT in order to obtain their galaxy groups}\label{tab:AAT_SNe}
\begin{threeparttable}
\begin{tabularx}{\textwidth}{l@{\extracolsep{\fill}}ccccccc}
\hline \hline
SN & Host & RA$_\textrm{Host}$ & DEC$_\textrm{Host}$ & $z_\textrm{hel}$\tnote{a} & $z_\textrm{grp}$\tnote{b} & $\mu$\tnote{c} & N$_\textrm{grp}$ \\
 & & (\degree) & (\degree) & & & (mag) & \\
\hline
1990T & WISEA J195859.25-561527.7\tnote{d} & 299.74704 & -56.25794 & 0.04017 & 0.04048 & -- & 4 \\
1991ag & IC 4919 & 300.03771 & -55.37339 & 0.01418 & 0.01599 & -- & 65 \\
1992P & IC 3690 & 190.70496 & 10.35747 & 0.02526 & 0.02452 & 35.4152 & 7 \\
1992ag & ESO 508- G 067 & 201.04383 & -23.87736 & 0.02484 & 0.02455 & 34.8529 & 3 \\
1994T & CGCG 016-058 & 200.37712 & -2.14556 & 0.03463 & 0.03482 & 36.009 & 2 \\
1999ac & NGC 6063 & 241.80412 & 7.97900 & 0.00951 & 0.00988 & 33.0149, 33.0668 & 4 \\
1999aw & SCP J110136.37-060631.6 & 165.40154 & -6.10878 & -- & -- & 36.3089 & -- \\
1999cp & NGC 5468\tnote{e} & 211.64537 & -5.45311 & 0.00938 & 0.00914 & 33.1428, 33.3164 & 17 \\
2000bh & ESO 573-14 & 185.31587 & -21.99575 & 0.02289 & -- & 34.9681 & -- \\
2002cr & NGC 5468\tnote{e} & 211.64537 & -5.45311 & 0.00938 & 0.00914 & 33.3316, 33.346 & 17 \\
2005bg & KUG 1214+166 & 184.32179 & 16.37150 & 0.02303 & 0.02292 & 35.1105 & 13 \\
2005ki & NGC 3332 & 160.11821 & 9.18256 & 0.01949 & 0.01937 & 34.5867, 34.6646, 34.6698 & 9 \\
2006ax & NGC 3663 & 170.99962 & -12.29644 & 0.01669 & 0.01657 & 34.4206, 34.4235 & 13 \\
2007ai & ESO 584- G 007 & 243.22233 & -21.62333 & 0.03166 & 0.03192 & 35.9491, 35.9488 & 8 \\
2007ca & MCG -02-34-061 & 202.76600 & -15.10119 & 0.01392 & 0.01420 & 34.4241, 34.2688, 34.3368 & 4 \\
2007cb & ESO 510- G 031 & 209.57150 & -23.37164 & 0.03642 & 0.03672 & -- & 64 \\
2007cc & ESO 578- G 026 & 212.17492 & -21.59725 & 0.02905 & 0.02969 & -- & 25 \\
2007cf & CGCG 077-100 & 230.78087 & 8.52839 & 0.03275 & 0.03460 & -- & 122 \\
2007cg & ESO 508- G 075\tnote{d} & 201.38904 & -24.65239 & 0.03320 & 0.03236 & -- & 24 \\
2008ar & IC 3284 & 186.15650 & 10.83903 & 0.02625 & 0.02568 & 35.5377, 35.3892, 35.4597 & 34 \\
2009aa & ESO 570- G 020 & 170.92067 & -22.27050 & 0.02728 & -- & 35.4192 & -- \\
2009ds & NGC 3905 & 177.27046 & -9.72983 & 0.01907 & 0.01895 & 34.603, 34.8001 & 7 \\
2009gf & NGC 5525 & 213.91346 & 14.28261 & 0.01842 & 0.01860 & 34.6714, 34.5823 & 3 \\
2010gp & NGC 6240\tnote{e} & 253.24529 & 2.40092 & 0.02448 & 0.02422 & 34.5571 & 7 \\
PS1-14xw & NGC 6240\tnote{e} & 253.24529 & 2.40092 & 0.02448 & 0.02422 & 35.2077 & 7 \\
PS15aii & LEDA 42943\tnote{d} & 191.19038 & 9.75726 & 0.04636 & 0.04600 & 36.694 & 14 \\
\hline
\end{tabularx} 
\begin{tablenotes}
\item[a] {A reliable redshift for the host galaxy for SN~1999aw was unable to be extracted from the data obtained from the AAT.}
\item[b] {The host galaxies for SN~2000bh and SN~2009aa were not deemed to be in a galaxy group.} 
\item[c] {Distance modulus values are from Pantheon+. No distance modulus is reported by Pantheon+ if the SN LC did not pass quality cuts. Multiple distance modulus values are given if multiple independent surveys provide LCs.}
\item[d] {Untargeted SN host galaxy.}
\item[e] {Both NGC~5468 and NGC~6240 have hosted multiple SNe (SN siblings).}
\end{tablenotes}
\end{threeparttable}
\end{table*}

\section{Data}\label{sec:data}
We present new data with which we look to define new galaxy groups for a sample of 23 SN host galaxies without galaxy-group information available in the literature.
The data analyzed here make use of the SN distances reported by the Pantheon+ analysis \citep{Brout22,Scolnic22}.

\subsection{AAT Data}
We obtain redshifts of nearby galaxies for SN host galaxies with undefined galaxy groups across 21 pointings of the Anglo-Australian Telescope (AAT).\footnote{\url{https://aat.anu.edu.au/}.}
The AAT is a 3.9-m telescope located at Siding Spring Observatory in New South Wales, Australia, equipped with several instrumentation options. 
We observed using the AAOmega spectrograph fed by the 392-fiber Two Degree Field (2dF) robotic fiber positioner front-end \citep{Sharp06}.\footnote{\url{https://aat.anu.edu.au/science/instruments/current/AAOmega}.}
Each pointing with 2dF was centered on a SN host galaxy at low redshift without a previously defined galaxy group.
All SN host galaxies selected have declinations between $-60\degree<\mathrm{DEC}<+15\degree$.
We provide general information for all SNe observed in our sample from the AAT in Table~\ref{tab:AAT_SNe}.
Each SN host galaxy is provided along with the host galaxy's coordinates, individual heliocentric redshift obtained by the AAT, group-averaged redshift from using a modified FoF algorithm (described in Section~\ref{subsubsec:AAT_groups}), all distance modulus values reported by Pantheon+, and the number of galaxies deemed to be in its group. 
We note that six SNe do not have a Pantheon+ distance modulus as the targets were selected before the Pantheon+ analysis was finalized, and the light curves (LCs) of these six SNe did not pass quality cuts.\footnote{Five out of six of these SNe are found to be in large groups of 24 or more members. Although it may be hypothesized that these large groups could cause the distance modulus values to be cut as outliers in the Pantheon+ analysis, instead, the majority of them get cut because they do not have data within five days of the estimated peak brightness.}

Of the 21 pointings, two of the targeted host galaxies have hosted SN siblings \citep[e.g.,][]{Kelsey24,Dwomoh24}, NGC~5468 hosting both SN~1999cp and SN~2002cr as well as NGC~6240 hosting both SN~2010gp and SN~PS1-14xw.
These siblings provide additional independent distance measurements which we use in our analysis.
An additional three SN host galaxies that we did not intentionally target were found in the fields we observed; these were WISEA J195859.25-561527.7 (SN~1990T), ESO~508- G~075 (SN~2007cg), and LEDA 42943 (SN~PS15aii).
Unfortunately, a reliable redshift for the host galaxy SCP~J110136.37-060631.6 was unable to be recovered from the spectrum we obtained with the AAT. 
This brings our sample of unique SN host galaxies observed with the AAT to 23 (that could be successfully assigned a redshift).

\begin{figure}
    \centering
    \includegraphics[width=\columnwidth]{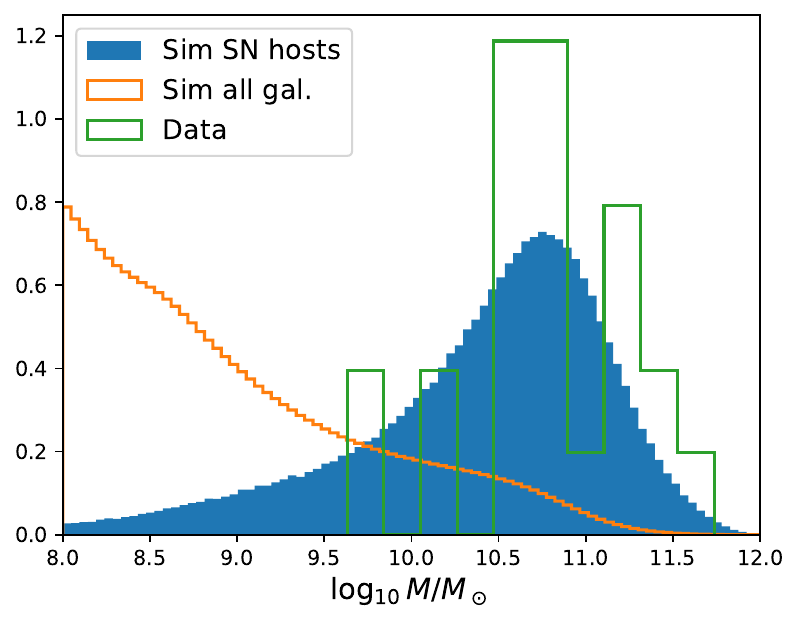}
    \caption{Normalized stellar mass distribution of the simulated Uchuu galaxy catalog (orange) compared to that from the simulated SN Ia host subsample (blue) and the data sample from the AAT (green).}
    \label{fig:massdist}
\end{figure}

Host galaxy masses for the data sample from the AAT are depicted in Fig.~\ref{fig:massdist} and calculated following \citet{Taylor11} by obtaining optical $g$- and $i$-band photometry from either the Sloan Digital Sky Survey \citep[SDSS;][]{SDSSDR18} or Pan-STARRS1 \citep[PS1;][]{Chambers16,Flewelling20} and the process described in \citet{Peterson24}.

\subsection{Pantheon+ Distances}
Distances come from the Pantheon+ analysis \citep{Brout22,Scolnic22}.
The sample analyzed by Pantheon+ is a compilation of 1701 SN Ia LCs from 18 different surveys with redshifts spanning 0.001--2.261.
SN Ia LCs in the Pantheon+ sample are fit with the SALT2 model \citep{Guy07,BroutSALT2} in order to obtain LC parameters such as LC stretch, $x_1$, and LC color, $c$, as well as the overall amplitude, $x_0$, which relates to the apparent magnitude, $m_B=-2.5\log(x_0)$.
Nuisance parameters such as $\alpha$ and $\beta$, which when multiplied by $x_1$ and $c$, respectively, become luminosity corrections, are globally fit for, and distance moduli, $\mu$, are then calculated using a modified Tripp relation \citep{Tripp98},

\begin{equation}
    \mu = m_B + \alpha x_1 - \beta c - \mathcal{M} - \delta_{\mu - \textrm{bias}},
\end{equation}

\noindent along with the fiducial SN Ia absolute magnitude, $\mathcal{M}$, and the calculated bias corrections, $\delta_{\mu - \textrm{bias}}$, to account for selection effects (see section 2.1 of \citet{Scolnic22} for details).
Ten SNe have multiple Pantheon+ distance modulus values listed in Table~\ref{tab:AAT_SNe} because the SNe were observed by multiple sources/surveys and therefore have multiple LCs from which distances are fit.

\begin{figure*}[!t]
    \centering
    \includegraphics[width=\textwidth]{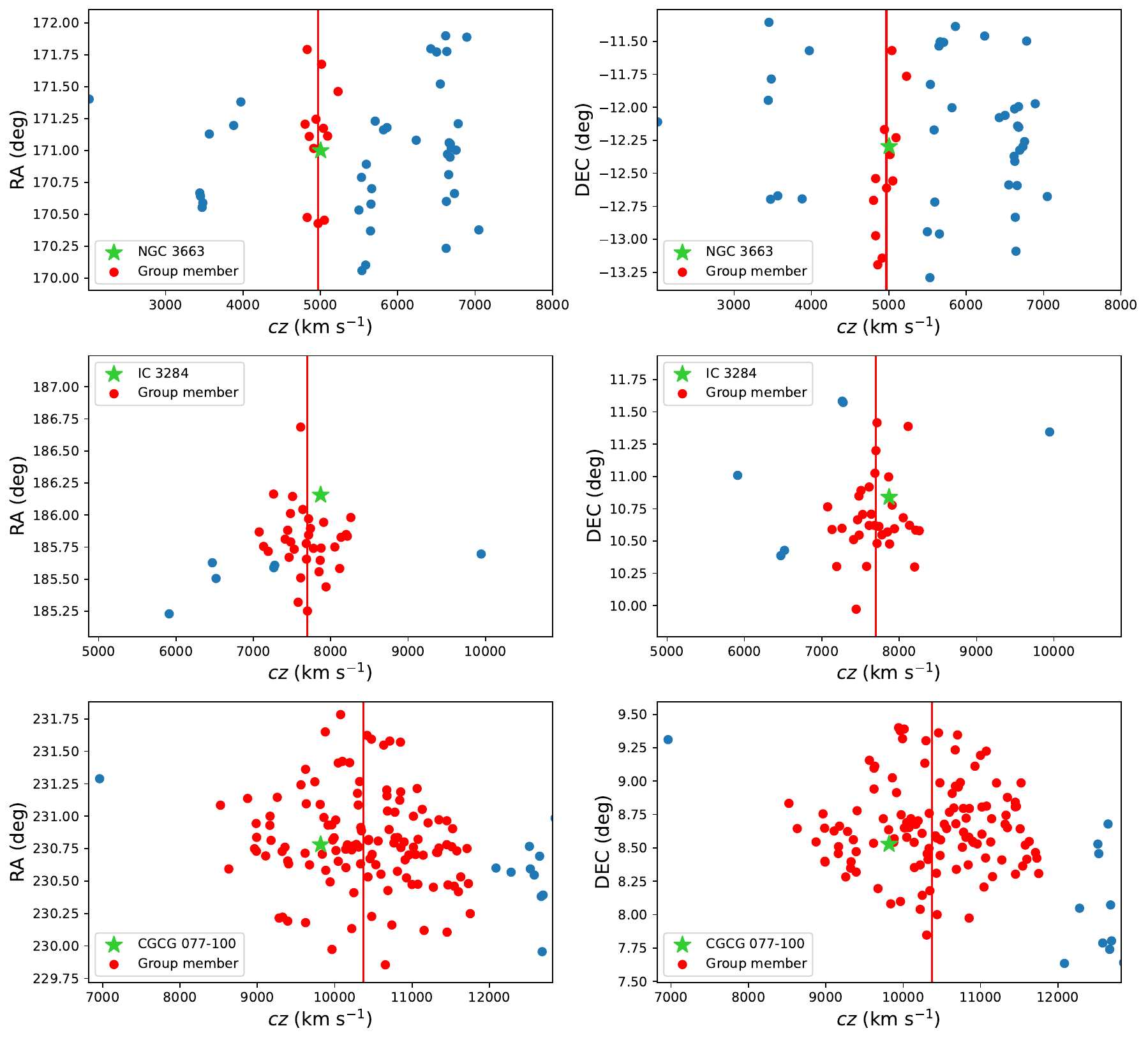}
    \caption{Example galaxy groups from the data for three pointings on the AAT. 
    The left and right panels provide RA vs.~\textit{cz} and DEC vs.~\textit{cz}, respectively, and each row references the same target galaxy. 
    Each panel spans 0.02 in redshift and is centered on the target galaxy (indicated with a green star and labeled in the legend). 
    Group members are indicated in red, while galaxies not defined in the targeted galaxy group are in blue. The red vertical line marks the group's average redshift.}
    \label{fig:example_coord_vs_vel_groups}
\end{figure*}

\section{Galaxy Groups}\label{sec:galaxygroups}
\subsection{Galaxy Groups in Simulations}\label{subsec:simulations}
To predict the number of SN host galaxies in groups and the improvement of Hubble residual scatter with group assignments, we use N-body simulations and identify a subsample of SN host galaxies following SN host characteristics presented in the literature.
For the N-body simulations, we make use of the Uchuu Universe-Machine (UM) simulated galaxy catalog \citep{Ishiyama2021,Aung23} which follows \citet{Behroozi19} who fit an empirical model to find the relationship between galaxies and halo properties in order for them to be in agreement with data.
The Uchuu UM catalog is based on an N-body simulation box with a side length of 2~\hGpc{}. 
Galaxies are generated with masses down to $5 \times 10^8$ \textit{M}$_\odot$ along with their properties such as stellar mass and star formation rate. 
The box is available as a snapshot at different redshifts; here we choose to use the one at $z=0$. The fiducial cosmology used in Uchuu is from \cite{Planck15}.
We split the main box into 64 non-overlapping sub-boxes of a volume corresponding to a redshift limit of $z\sim0.085$.

In order to define the galaxy groups in our sub-boxes, we use a simple FoF algorithm with this general process: (i) choose a galaxy in the catalog, (ii) find all neighbors within a sphere of radius $l$ and add them to the group, (iii) iterate over step two for all neighbors until no new neighbors are found, and (iv) iterate from step one until all galaxies have been assigned. 
We consider pairs of galaxies to be in groups in this work, and as a baseline we choose a value of $l = 0.3$~\hMpc{} which is similar to linking lengths/distances used in other works \citep[i.e.,][]{Tully15,Lambert20}. 
To reduce computation time, the FoF algorithm is run only for galaxies with a redshift $z < 0.055$ which comfortably encompasses all the SNe in our sample.

From the Uchuu galaxy catalog we select a subsample of 500,000 galaxies across the 64 mocks that corresponds to SN Ia host galaxies. This selection is made by randomly drawing hosts according to the stellar mass distribution of SN Ia hosts as described in \citet{Wiseman21} and the star formation rate (SFR) distribution of SN Ia hosts as described in \citet{Sullivan06}.\footnote{Specifically, the ``A+B'' model from \citet{Sullivan06}.} In Fig.~\ref{fig:massdist} we show the stellar mass distribution of all Uchuu galaxies compared to that of our SN Ia hosts subsample.
We additionally include the mass distribution of host galaxies in our data sample from the AAT.
As expected, the masses from the data tend to be more massive and are more consistent with the simulated SN host distribution than the complete distribution of simulated galaxies.

\begin{figure*}
    \centering
    \includegraphics[width=\linewidth]{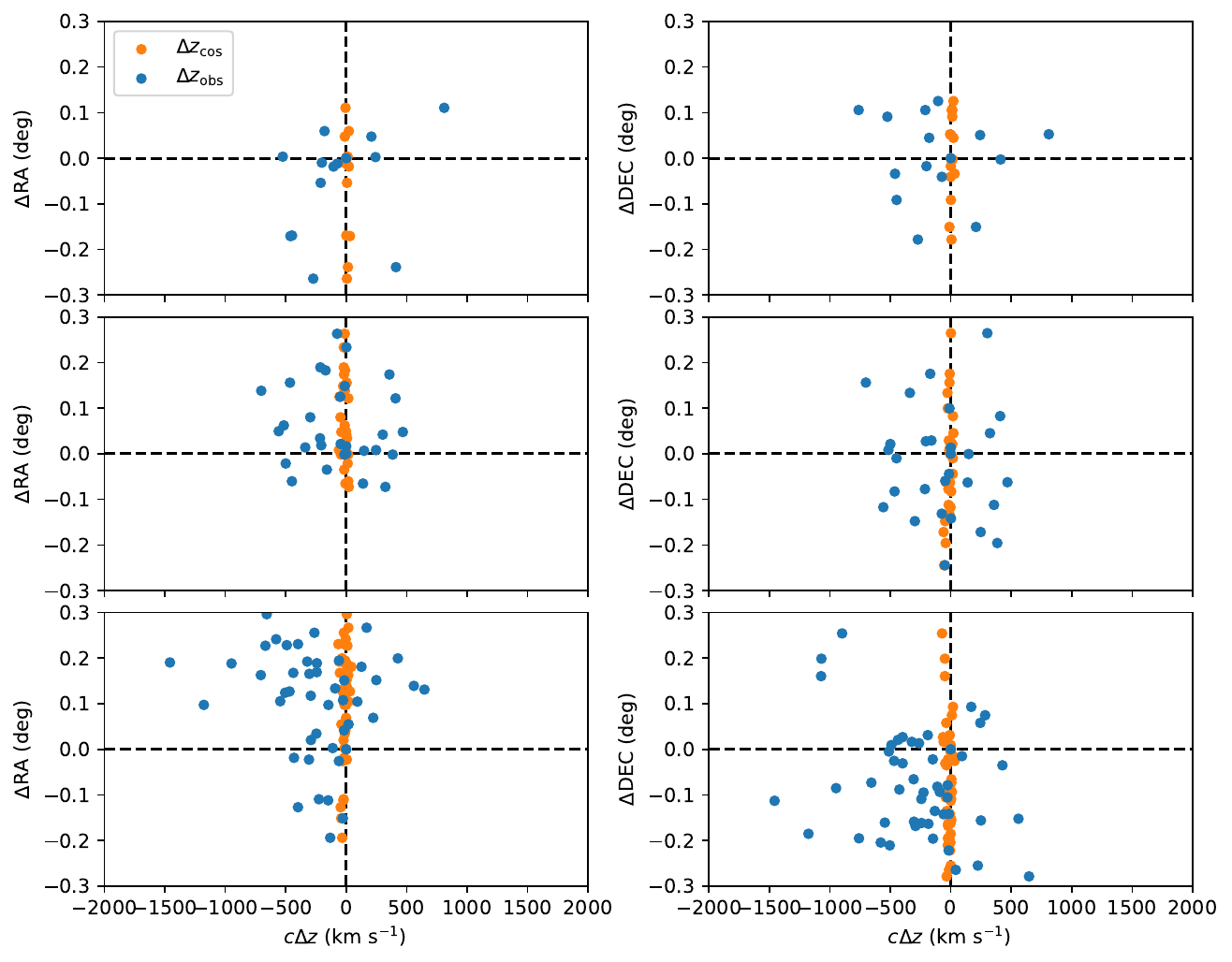}
    \caption{Similar to Fig.~\ref{fig:example_coord_vs_vel_groups}, but example galaxy groups from the Uchuu simulations in each row and relative coordinates and velocities on each axis. Observed redshifts (calculated from velocity and redshift information provided by the Uchuu catalog) are indicated in blue and true cosmological redshifts are in orange.}
    \label{fig:sim_coord_vs_vel_groups}
\end{figure*}

\subsection{Galaxy Groups in Data}\label{subsec:group_defs}

For the sample we obtain from the AAT, we define galaxy groups using a modified FoF algorithm \citep{Lambert20}.
To increase our sample size, we also include groups defined by T15 already identified in Pantheon+ \citep{Peterson22} in this analysis.

\subsubsection{Group Definitions with AAT Data}\label{subsubsec:AAT_groups}

With the data from the AAT, we employ a modified FoF algorithm, which was used in \citet{Lambert20} for the 2MRS galaxy-group catalog, to define our galaxy groups.\footnote{\url{https://pypi.org/project/fofpy/}.}
This modified FoF group finder is based on graph theory.
This algorithm is more complicated than what is used for the simulations since the simulations do not suffer from magnitude limitations and can use real space coordinates.
The one modification we make to the original FoF algorithm from \citet{Lambert20} is that groups with two members are included rather than a minimum of three members.
Although \citet{Lambert20} argue for a minimum group size of at least three members, works such as \citet{Tully15}, \citet{Lim17}, and \citet{Peterson22} include binaries in their definition of groups. 
We opt to include binaries in this analysis in order to increase our sample size. 

Group sizes for this sample from the AAT are reported in Table~\ref{tab:AAT_SNe}, and the median group size is 9 galaxies. 
Of the SN host galaxies in our sample with a reliable redshift from the AAT, two of them were not found to be in a group --- ESO 573-14 (SN~2000bh) and ESO 570- G~020 (SN~2009aa).
Thus, 91\% (21/23) of our sample of unique SN host galaxies are found to be in groups.

\newcommand{\mywidth}{0.49}
\begin{figure*}[!t]
    \includegraphics[width=\mywidth\textwidth]{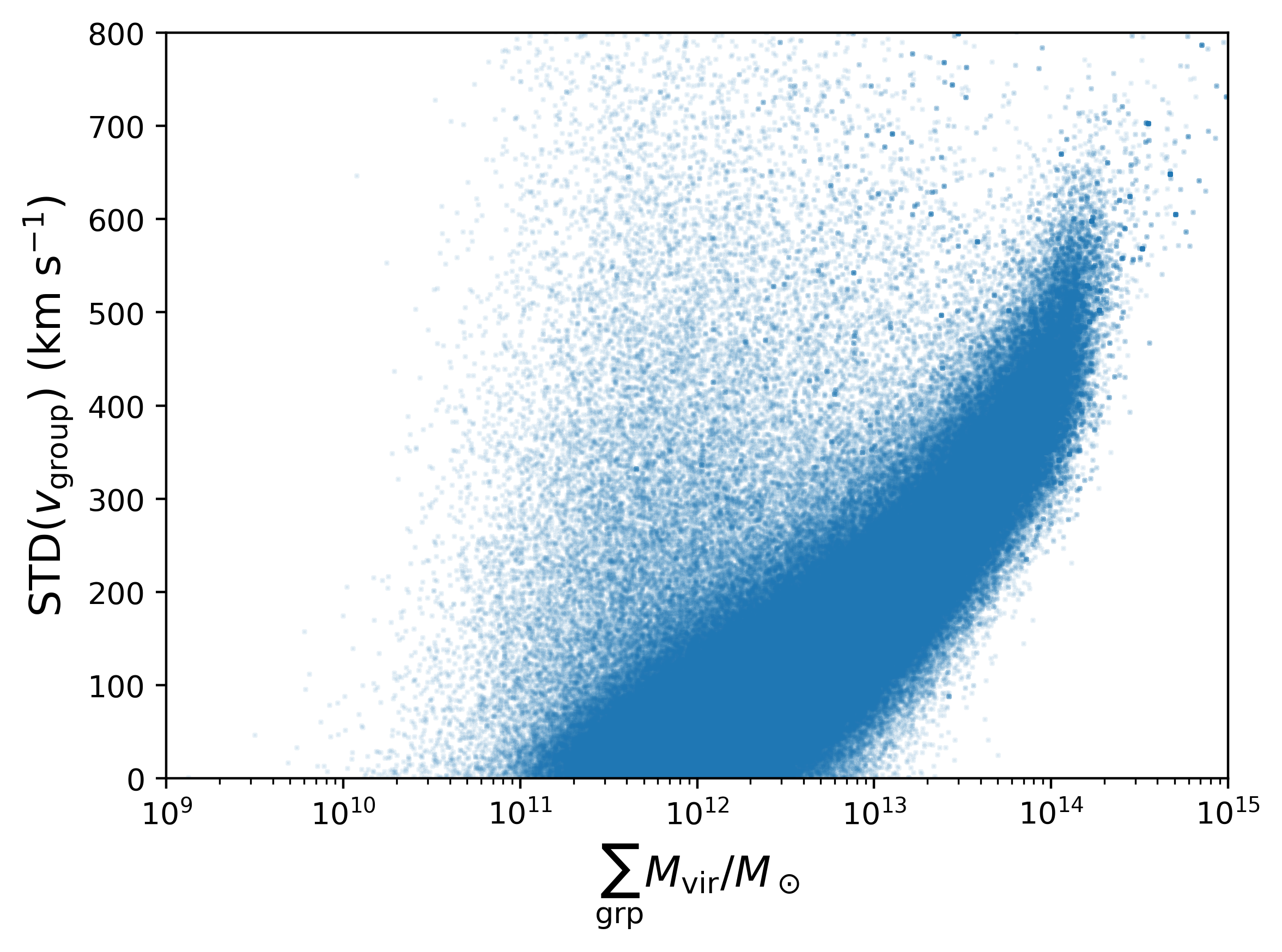}
    \includegraphics[width=\mywidth\textwidth]{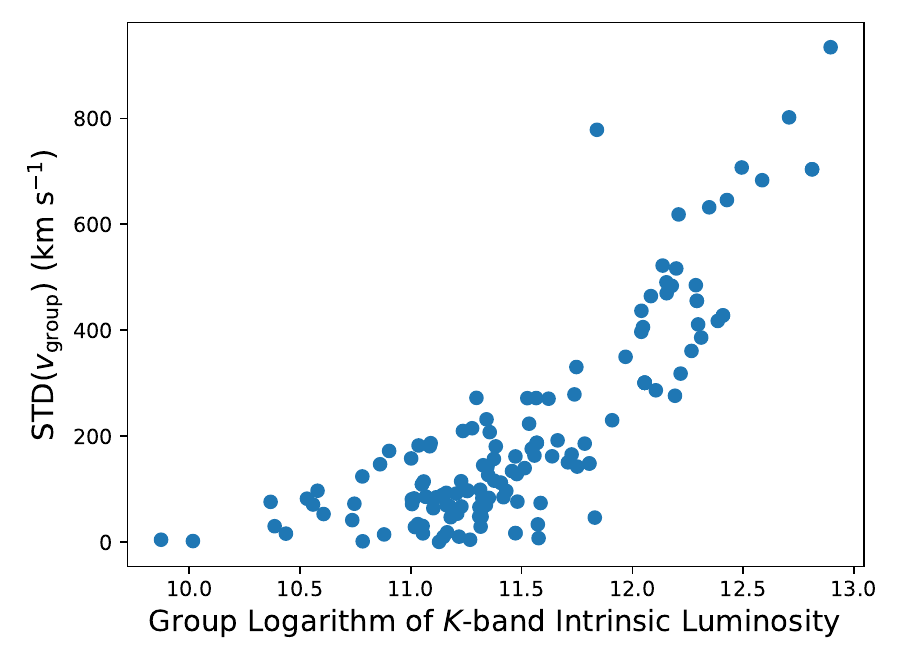}
    \caption{\textbf{Left}: Standard deviation of velocities of simulated galaxies inside groups with respect to the virial masses of groups for the SN host distribution from Uchuu. \textbf{Right}: Standard deviation of velocities of Pantheon+ galaxies inside groups with respect to the group logarithm of $K$-band luminosity as a proxy for the galaxy group's mass.}\label{fig:std_vs_mass}
\end{figure*}

Three example groups from our data are depicted in Fig.~\ref{fig:example_coord_vs_vel_groups} for the host galaxies NGC~3663 (SN~2006ax), IC~3284 (SN~2008ar), and CGCG~077-100 (SN~2007cf).
Each row corresponds to a single target host galaxy, where the left panel is right ascension versus \textit{cz}, and the right panel is declination versus \textit{cz}.
Galaxy-group members are indicated in red, and each group's average redshift, which is obtained by taking the mean of all redshifts in the group,\footnote{\citet{Peterson22} explored a weighted mean for group-averaged redshifts and did not find the results to be significantly different than unweighted averaging.} is marked by a red vertical line.
The galaxy group defined for CGCG~077-100 (Fig.~\ref{fig:example_coord_vs_vel_groups} lower panels), with 122 members, is the largest galaxy group defined in this sample, and the difference between the target galaxy's redshift and the group-averaged redshift is 554 km~s$^{-1}$.
Coordinates versus velocity plots for the complete sample are provided in Appendix~\ref{appendix:all_ra_dec_vs_velocity}.

\subsubsection{Groups in Pantheon+ Data}\label{subsubsec:T15_groups}
We supplement the galaxy groups identified using our AAT data with galaxy groups already defined by T15 in Pantheon+ \citep{Peterson22}.
The T15 group catalog makes use of redshifts from 2MRS \citep{Huchra12}\footnote{Complete out to $K_s \approx 11.75$ mag.} and an iterative process based on velocity and distance associations in order to construct groups. 
From the Pantheon+ data release \citep{Scolnic22,Carr21}, there are 143 SNe Ia with a group defined in T15 as well as a reported distance modulus in Pantheon+.

\section{Galaxy Group Results}\label{sec:GGresults}
\subsection{Comparing Simulations with Data}
Three example galaxy groups from the simulations are provided in Fig.~\ref{fig:sim_coord_vs_vel_groups}.
This figure is similar to Fig.~\ref{fig:example_coord_vs_vel_groups}, but here we provide coordinates and velocities relative to the assigned SN Ia host galaxy and also indicate the true cosmological redshifts as compared to the observed redshifts.
Visually, the groups observed in the data are comparable to the groups observed in the simulations.

From the simulations we obtain a median difference between the individual redshifts and the group redshifts of 2.1 km~s$^{-1}$ which is near zero, as expected, and a median absolute difference of 68.3 km~s$^{-1}$. 
The median difference for the combined data is 13.7 km~s$^{-1}$, while the median absolute difference is 105.0 km~s$^{-1}$.
Considering the standard error of the differences between the individual redshifts and group redshifts in the data of 21.6 km~s$^{-1}$, these values are consistent with those from the simulations.

With the simulations, we plot the standard deviation (STD) of velocities in groups for SN hosts as a function of the group's virial mass in the left panel of Fig.~\ref{fig:std_vs_mass}.
Larger mass groups tend to have more dispersion in their velocities.
Despite this, there are low-mass groups that have high dispersion as well.
We deem these groups to be ``unvirialized."
The median and mean STD of velocities of simulated groups here are 135 km~s$^{-1}$ and 175 km~s$^{-1}$, respectively.
We can expect a general improvement to our redshifts of this order from using groups.

For the Pantheon+ data, we provide the STD of velocities in groups versus group $K$-band intrinsic luminosity as a proxy for mass in the right panel of Fig.~\ref{fig:std_vs_mass}.
Both $K$-band luminosities and velocities of galaxies in groups are provided by T15.
We observe a general trend of increased STD as a function of $K$-band luminosity similar to what is seen in the left panel of Fig.~\ref{fig:std_vs_mass}, but we do not find ``unvirialized" low-mass, large-STD groups in the data.
This could be due to the magnitude-limited factor of the galaxy-group catalog, where low-mass galaxies go undetected and thus do not contribute to velocity scatter.
The median STD of velocities of galaxies in groups in Pantheon+ is 145 km~s$^{-1}$, the mean is 208 km~s$^{-1}$, and the standard error is 16 km~s$^{-1}$. 
These values are roughly consistent with, albeit slightly higher than, those observed in the simulations.

\begin{figure}[!bt]
    \centering
    \includegraphics[width=\columnwidth]{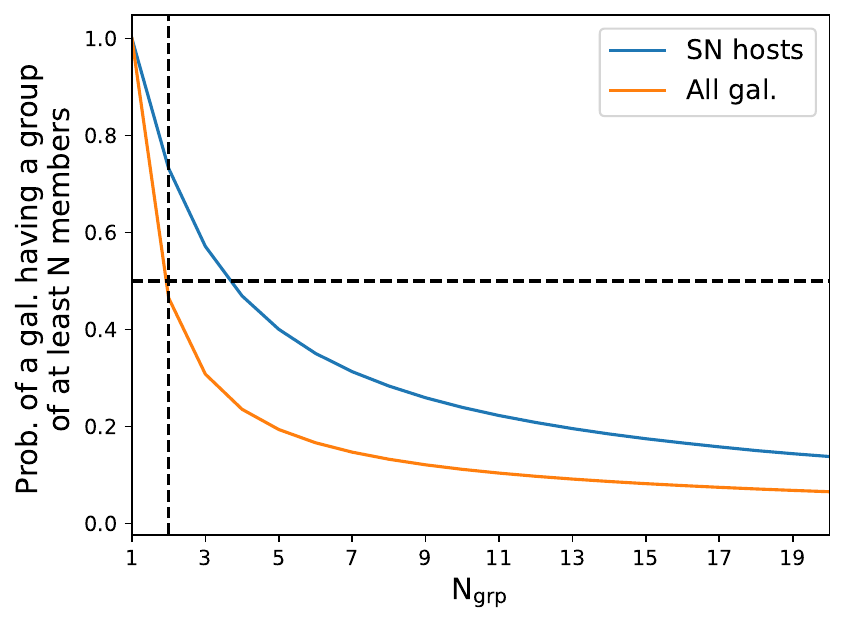}
    \caption{Probability of a given galaxy of having at least N members in its galaxy group in the simulations. Two distributions are provided: the complete distribution of galaxies in orange and the subset of galaxies that host SNe in blue. The vertical line marks two members in a group, while the horizontal line marks 50\% probability.}
    \label{fig:sim_how_many_in_groups}
\end{figure}

\subsection{Frequency of Galaxies in Groups}\label{subsec:freq_groups}

As stated in Section \ref{subsubsec:AAT_groups}, 91\% of the data sample from the AAT are found to be in groups.
In Fig.~\ref{fig:sim_how_many_in_groups}, we present the probability distributions of a given galaxy in our simulations of being part of a group of at least N members.
According to the simulations, 73\% of SN host galaxies should be found in a group of at least two members, while 47\% of the full distribution of galaxies should be found to be in groups.
These values have been provided in Table~\ref{tab:percent_in_groups} along with the percentages of galaxies in groups reported in the literature.
From previous works \citep[i.e.,][]{Crook07,Tully15,Peterson22}, SN host galaxies reportedly have a lower percentage of galaxies in groups, while in our simulations, the opposite is true.
We believe this is due to the magnitude-limited nature of the galaxy catalogs analyzed.

The difference in percentages for SN hosts between what was found in \citet{Peterson22} (30\%), the simulations in this work (73\%), and the data sample in this work (91\%) is also worth noting.
The 30\% reported in \citet{Peterson22} may be lower than the true percentage of SN host galaxies in groups given that a large portion ($\sim$47\%) of the host galaxies themselves could not be identified in the galaxy catalogs used for group identification.
For a measure of uncertainty on the percentage of SN hosts found in groups in our data, we bootstrap our simulations by obtaining the same sample size and same redshift distribution as our data and calculating the percentage in groups 100 times, and we take the standard deviation of the set.
With this method, we obtain an estimate for uncertainty on the percentage from the data of 8\%.
The percentage of SN host galaxies in groups in the data then is approximately 2.3$\sigma$ larger than the prediction from simulations.
We note that the simulations do not have a magnitude limitation, and therefore, the simulated sample may include a larger percentage of low-mass/dim galaxies than the data sample, and we also recognize the limited sample size of this pilot program (at 23 host galaxies). Each of these reasons could cause the difference in percentages to be larger than expected, which we leave for further analysis.

\begin{table}[!t]
\caption{Percent of galaxies found to be in groups for different works}\label{tab:percent_in_groups}
\begin{tabularx}{\columnwidth}{l@{\extracolsep{\fill}}cr}
\hline \hline
Work & \% in Groups & Notes \\
\hline
\multirow{2}{*}{\citet{Crook07}} & \multirow{2}{*}{73\%} & \multirow{2}{*}{All gal., 2MRS}\\
(data) & & \citep{Huchra05a}\\
\multirow{2}{*}{\citet{Tully15}} & \multirow{2}{*}{58\%} & \multirow{2}{*}{All gal., 2MRS}\\
(data) & & \citep{Huchra12}\\
\multirow{2}{*}{\citet{Peterson22}} & \multirow{2}{*}{30\%} & \multirow{2}{*}{SN hosts, Pantheon+}\\
(data) & & \\
\hline
\multirow{2}{*}{This work (sims)} & \multirow{2}{*}{47\%} & \multirow{2}{*}{All gal., Uchuu}\\
& & simulations\\
\multirow{2}{*}{This work (sims)} & \multirow{2}{*}{73\%} & \multirow{2}{*}{SN hosts, Uchuu}\\
& & simulations\\
\hline
\multirow{2}{*}{This work (data)} & \multirow{2}{*}{91\%} & \multirow{2}{*}{Targeted SN hosts}\\
& & on the AAT\\
\hline
\end{tabularx}
\end{table}

\section{Impact on Hubble Residuals from Groups}\label{sec:HRresults}

Using the full data sample of 173 SNe from the AAT and Pantheon+ as well as simulations, we evaluate the improvement from using group-averaged redshifts in two separate ways. The first is from analyzing how often Hubble residuals are improved when using the group redshift, and the second is from measuring how much the Hubble residual scatter improves for different bins of group sizes and redshifts.

\subsection{Frequency of Improved Hubble Residuals with Groups}

\begin{figure}[!t]
    \centering
    \includegraphics[width=\columnwidth]{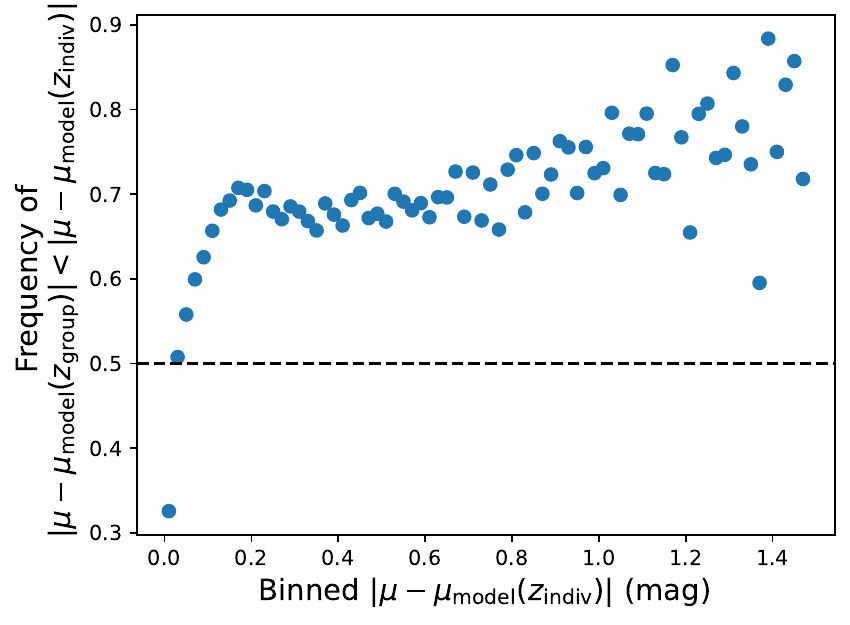}
    \caption{Frequency at which the group redshift results in an improved Hubble residual for bins of absolute Hubble residuals for the simulations. A dotted horizontal line at a frequency of 0.5 is provided.}
    \label{fig:sim_percent_by_mures_bin}
\end{figure}

\begin{figure}[!t]
    \centering
    \includegraphics[width=\columnwidth]{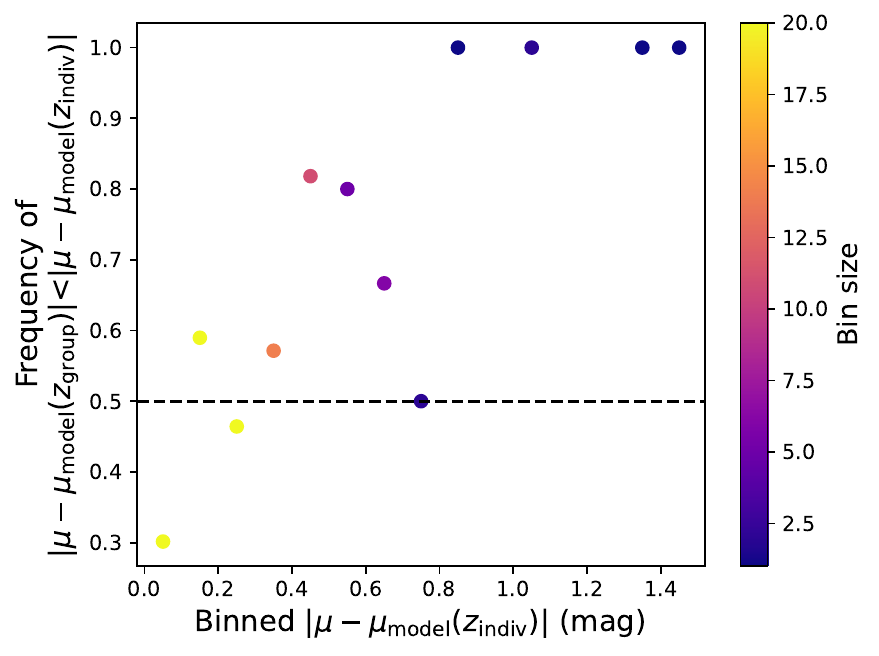}
    \caption{Frequency at which the group redshift results in an improved Hubble residual for bins of absolute Hubble residuals for the full data sample. Points are colored by the number of groups in the respective bin. A dotted horizontal line at a frequency of 0.5 is provided.}
    \label{fig:data_percent_by_mures_bin}
\end{figure}

\begin{figure*}[!t]
    \centering
    \includegraphics[width=\textwidth]{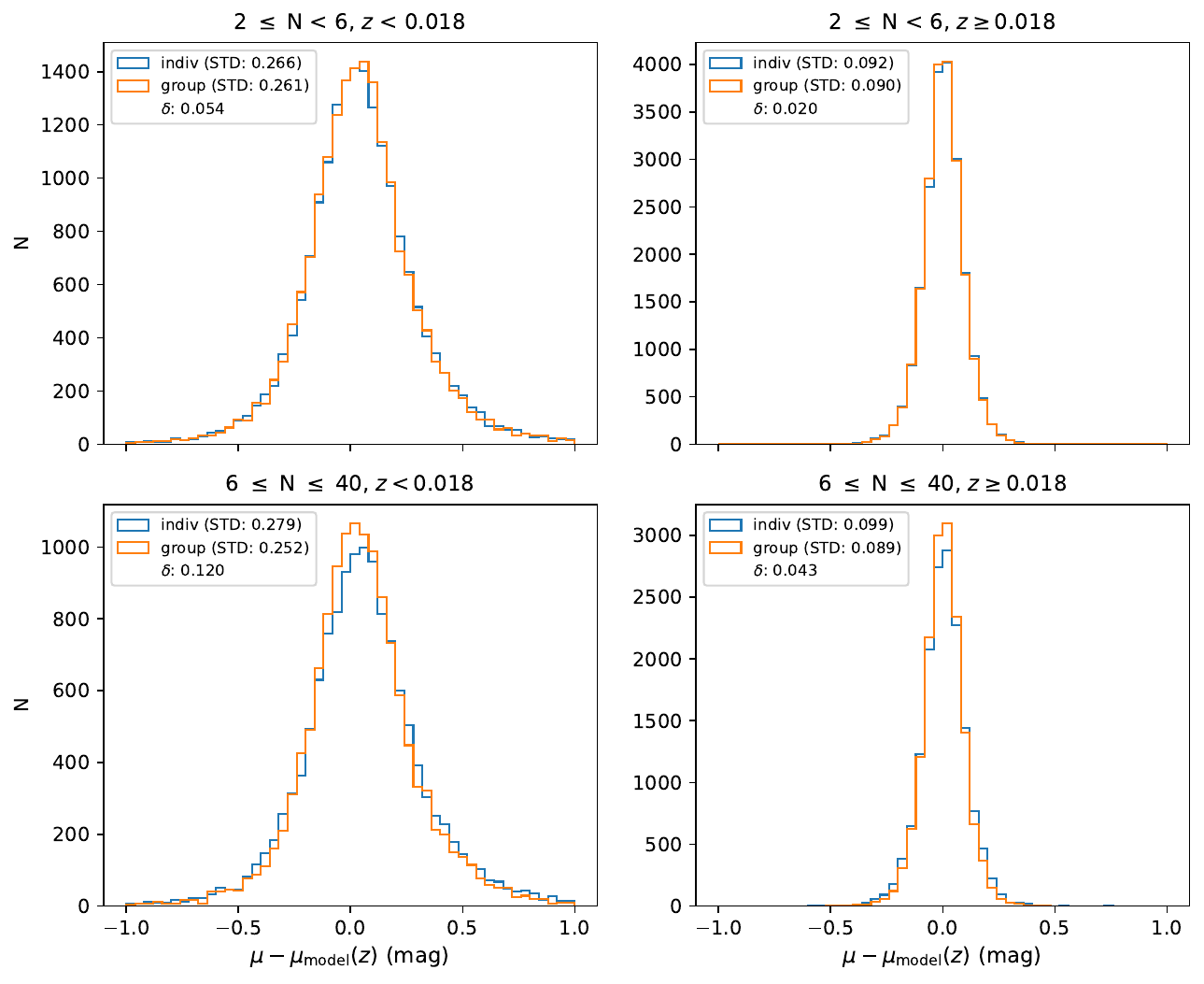}
    \caption{Binned Hubble residual distributions for individual redshifts (blue) and group-averaged redshifts (orange) with simulations. STD values for each distribution and the differences in STD values in quadrature, $\delta$, are provided in the legend. The specific bin descriptions are given in each panel's title.}
    \label{fig:binned_sim}
\end{figure*}

\begin{figure*}[!t]
    \centering
    \includegraphics[width=\textwidth]{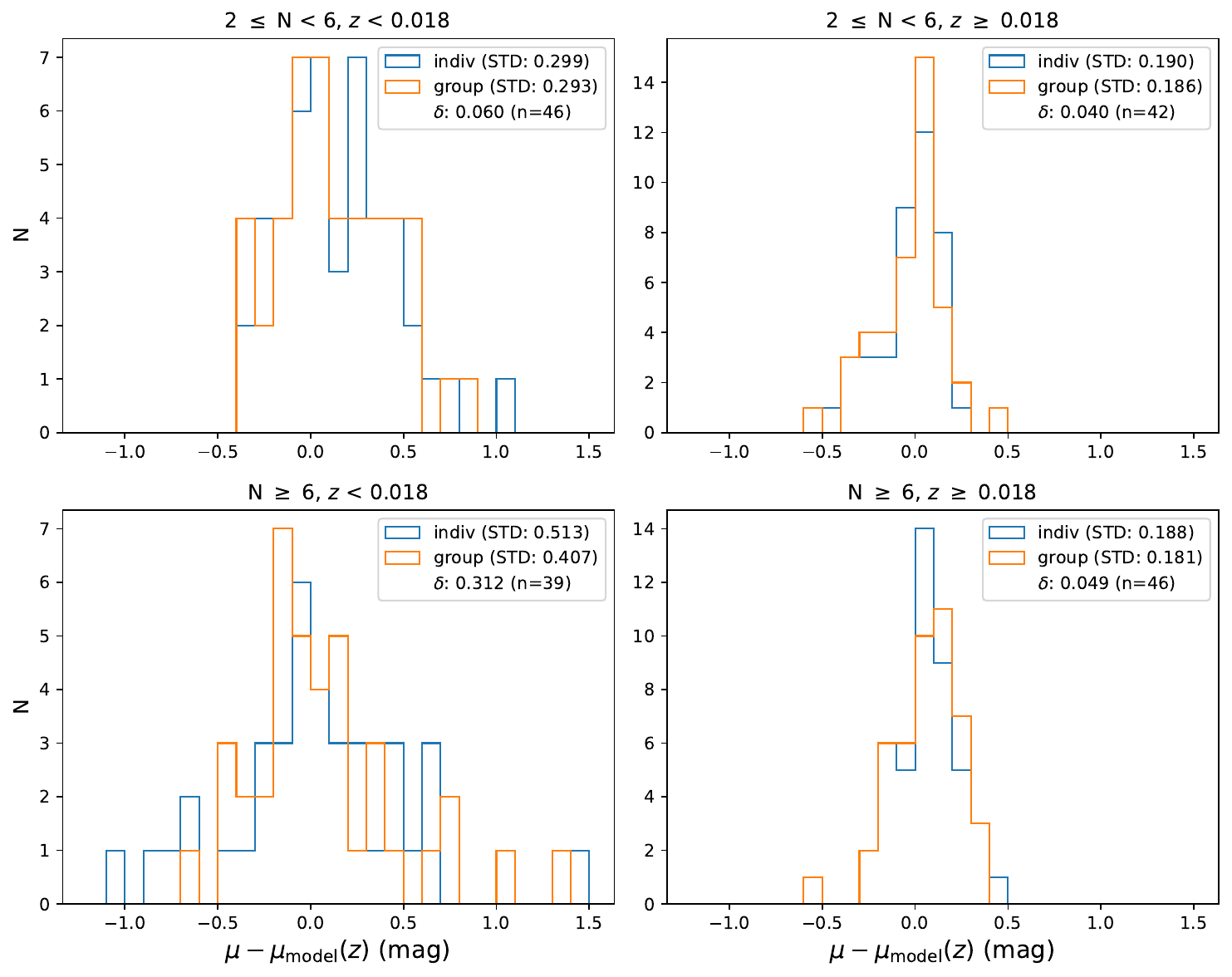}
    \caption{The same as Fig.~\ref{fig:binned_sim}, but with data. Binned Hubble residual distributions for both individual redshifts as well as group-averaged redshifts.
    The STD for each distribution is provided in the legend as well as the difference in scatter in quadrature ($\delta$) and the binned sample size. Bin edges were chosen to produce roughly equal samples sizes in each bin in terms of both redshift and group size with the low-redshift values in the left panels and small group sizes in the upper panels.} 
    \label{fig:binned_data_2x2}
\end{figure*}

One significant finding from our analysis is the fact that, in both the simulations and data, the magnitude of Hubble residuals is reduced when using group redshifts only about 50--55\% of the time.
This is likely due to the fact that not all PVs, such as larger-scale bulk motions, are accounted for when using group redshifts.
That being said, according to the data, using group redshifts results in large improvements in Hubble residuals ($>$0.1 mag) far more frequently (15\% of the time) than worsening the residual by a large amount (5\% of the time).
We further detail these results in Appendix~\ref{appendix:cumul_dists}.

In order to better understand this frequency of improved Hubble residuals from using groups, we plot this frequency as a function of the magnitude of the Hubble residuals when using the individual redshift in bins with the simulations in Fig.~\ref{fig:sim_percent_by_mures_bin}.
The prediction from simulations is that generally, the larger the Hubble residual is without galaxy group information, the more frequently grouping results in a reduction in the magnitude of the Hubble residual, up to $\gtrsim$70\% of the time for Hubble residuals $>$0.8 mag.

The frequency at which the group redshift results in a better Hubble residual as a function of the magnitude of the Hubble residual for the data is given in Fig.~\ref{fig:data_percent_by_mures_bin}.
In the data, it is even more apparent that the larger the magnitude of the Hubble residual is originally, the more often the group redshift improves the Hubble residual.
When the magnitude of the Hubble residual is $>$0.8 mag using the individual redshift, group-averaging reduces the size of the Hubble residual 100\% of the time.
This shows that large Hubble residuals previously seen at low redshift could be due to a lack of group assignments.

\subsection{Hubble Residual Dispersion with Galaxy Groups}

We present Hubble residual histograms in bins of redshift and group size for the simulations in Fig.~\ref{fig:binned_sim}.
In order to compare the Hubble residual scatter between the simulations and data, for Fig.~\ref{fig:binned_sim} we match the simulated redshift distribution to that found in the data. 
Residuals are provided for both individual and group redshifts.
Low-redshift panels are on the left ($z<0.018$), and small groups are in the upper panels ($\textrm{N}<6$).
STD values for these various bins of Hubble residuals are provided in the legends as well as an additional statistic,
\begin{equation}
    \delta=\sqrt{\textrm{STD}_\textrm{indiv}^2-\textrm{STD}_\textrm{group}^2},
\end{equation}
\noindent where $\delta$ is the difference in Hubble residual STDs in quadrature.
The largest STD values are predicted to come from the low-redshift bins.
This is expected, since PVs make up a larger percentage of the overall velocity at low redshift.
The group-averaged redshifts demonstrate improved Hubble residual scatter in all bins, and the largest improvement comes from large groups at low redshift with a $\delta$ of 0.120 mag.\footnote{In the case where the group scatter is larger than the individual scatter, the $\delta$ parameter is imaginary. From the same method of bootstrapping described in Section~\ref{subsec:freq_groups}, we estimate there is a 12\% chance of this happening for our sample size of 173 hosts.}

\begin{figure*}[!t]
    \centering
    \includegraphics[width=\textwidth]{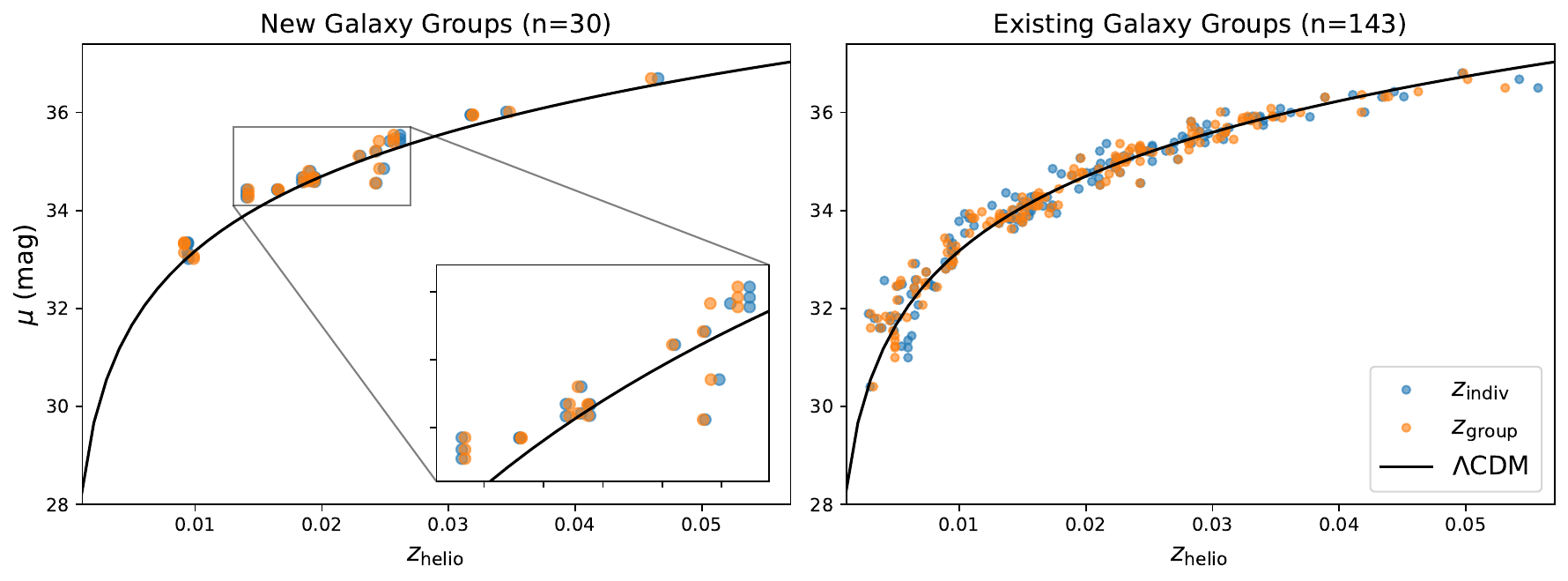}
    \caption{Hubble diagrams using both newly defined galaxy groups from the AAT data (left panel) and existing galaxy groups in the T15 catalog (right panel). Individual redshifts (blue) and group averaged redshifts (orange) are plotted in each Hubble diagram as well as a reference cosmology (black).}
    \label{fig:HD_new_old}
\end{figure*}

We provide Hubble residual distributions for the data for both individual redshifts as well as group-averaged redshifts binned by both redshift and group size in Fig.~\ref{fig:binned_data_2x2}.
Host galaxies in small groups with 2--5 members are plotted in the upper panels while larger groups with six or more members are given in the lower panels.
Higher redshift bins ($z\geq0.018$) are provided in the panels on the right.
Bin edges were selected to obtain roughly equal sample sizes ($\sim$40--45) in each bin for both low/high redshift and small/large group sizes.
STD values for each distribution are provided as well as the $\delta$ statistic, the difference in Hubble residual STDs in quadrature.
The panel with the largest improvement in Hubble residual scatter (from a STD$_\textrm{indiv}$ of 0.513 mag to a STD$_\textrm{group}$ of 0.407 mag) is the lower left panel for large group sizes at low redshift and a $\delta$ of 0.312 mag.
We note that both distributions in the lower left panel have large scatter values, but the improvement in scatter when using groups is the most apparent.
All bins result in improved Hubble residual scatter, with the upper right panel with small group sizes at higher redshift resulting in the smallest improvement ($\delta$ = 0.040 mag).
For the complete unbinned sample of 173 SNe, the STD value goes from 0.324 mag with individual redshifts to 0.285 mag with group redshifts.
This corresponds to a $\delta$ value of 0.155 mag. 
Following the same bootstrapping method described in Section~\ref{subsec:freq_groups}, from simulations we find the expected quartiles on our data's $\delta$ value for the unbinned sample to be 0.080 mag, 0.147 mag, and 0.190 mag at 25\%, 50\%, and 75\%, respectively.

The trends in $\delta$ values observed in the data are similar to those observed in the simulations. 
For both simulations and data, the largest $\delta$ value comes from large groups at low redshift, and the smallest $\delta$ value from small groups at higher redshift.
Further, the $\delta$ value from small groups at low redshift is slightly larger than the $\delta$ value from large groups at higher redshifts in both the simulations and data.

In the left panel of Fig.~\ref{fig:HD_new_old}, we present the Hubble diagram for all newly defined galaxy groups from the AAT data. 
Individual redshifts as well as group redshifts are provided as well as a fiducial cosmology (flat $\Lambda$CDM, $H_0=70$ km~s$^{-1}$~Mpc$^{-1}$, $\Omega_\textrm{m}=0.3$).
The Hubble diagram for all Pantheon+ SN host galaxies already found in groups is given in the right panel of Fig.~\ref{fig:HD_new_old}.
All distance modulus values are from Pantheon+, and for those SNe with multiple distance modulus values reported due to multiple surveys observing them, multiple points are plotted on the figure.
For SNe with larger residuals (larger than the median residual) initially, group-averaged redshifts result in points closer to the fiducial cosmology more frequently than the individual redshifts at a rate of 60\% and 67\% in the left and right panels, respectively.

\section{Discussion}\label{sec:discussions}

Our implementation of the Uchuu simulations have potential limitations in how well they can describe the data. For example, we select for SN host galaxies with a limited number of selection choices (mass and SFR), which may not describe the data fully. Further specific limitations and assumptions made for the Uchuu simulation framework are described in \citet{Behroozi19} and \citet{Aung23}.
When we select our SN host galaxies from the full Uchuu simulation on mass alone rather than mass and SFR, our results are modified only slightly, with the number of SN host galaxies found in groups in the simulations dropping by a few percent.
For another test, when we limit our SN hosts to $>$9.5~dex to even more closely match our mass distribution from the data, our conclusions remain the same.

\subsection{Potential Impact on Current and Future Surveys}
There are a number of recent surveys such as the Asteroid Terrestrial-impact Last Alert System \citep[ATLAS;][]{ATLAS,Smith20}, the Dark Energy Bedrock All-sky Supernova Survey (DEBASS, PI:~Brout), the Young Supernova Experiment \citep[YSE;][]{YSE}, and ZTF \citep{Bellm19} that have added a few hundred to a few thousand LCs to the complete sample of observed SNe Ia.
These surveys provide a quality sample set that can be used to search for many more undefined galaxy groups and improve measurements of cosmological redshifts on a larger scale than the one presented in this work.
We predict that of this more recent set of $\sim$5000 SNe, an estimated 30\% of them have preexisting galaxy-group information available in the literature, while new galaxy groups could be defined for $\sim$20--30\% more ($>$1000 SN hosts).
Even further, the \textit{Rubin} Observatory's Legacy Survey of Space and Time \citep[LSST;][]{LSST_DC2,Sanchez22} will begin imminently, and it is expected to observe thousands more SNe at low redshift for which galaxy groups could be defined. 

A separate follow-up survey targeting galaxy-group redshifts for SN hosts from the current samples as well as the future sample from LSST is an important and cost-effective way to improve the constraining power of these SNe Ia.  
Besides improving the statistical constraining power of each low-$z$ SN, improved redshifts would help separate the scatter due to peculiar velocities from the intrinsic scatter of the SNe. As a different or evolving intrinsic scatter between low- and high-redshift SNe is an important systematic uncertainty \citep{Brout19}, better measurements of the low-$z$ intrinsic scatter is a priority.

The Time-Domain Extragalactic Survey \citep[TiDES;][]{TiDES} is a follow-up spectroscopic survey of transients on the 4-metre Multi-Object Spectroscopic Telescope (4MOST) which has the goal of obtaining 30,000 live transient observations over the next five years.
TiDES is the only substantial spectroscopic survey currently aimed at classifying LSST transients, and although it is aiming to obtain on the order of 50,000 host galaxy redshifts, this is only $\sim$1--2\% of the full sample of SNe that will be observed with LSST.
Given this disparity, other spectroscopic instruments/surveys can and should target LSST transients as well as their galaxy groups. 
The Dark Energy Spectroscopic Instrument (DESI) LOW-$Z$ Secondary Target Survey \citep{DarraghFord23} is expected to be roughly complete out to $z=0.03$, but it will observe primarily in the Northern Hemisphere.
The AAT using 2dF is a perfect candidate for a galaxy-group targeted survey in the Southern Hemisphere. 
Other telescopes that can and should consider targeting SN host galaxy groups include the Wisconsin-Indiana-Yale-NOIRLab (WIYN) Telescope, which is equipped with the Hydra multi-object spectrograph and can observe $\sim$100 objects simultaneously within a one-degree radius. Subaru, which will soon be equipped with the Prime Focus Spectrograph (PFS), will be able to observe up to approximately 2400 objects at a time with a 1.3-degree field-of-view.
With the spectrographs available, not only can we spectroscopically confirm more LSST SNe but also obtain information on their galaxy groups.

Before LSST officially begins, there might be a dearth of transient candidates for observation on multi-object fiber spectrographs such as 4MOST. 
TiDES and/or other spectroscopic instruments/surveys have the opportunity to obtain more galaxy groups of both historical SNe in the literature as well as SNe that will continue to be observed by ongoing surveys such as ATLAS and ZTF.

\section{Conclusions}\label{sec:conclusions}

Using both simulations and data, we set out to answer two main questions: (i) how many galaxies are in groups, and specifically, how many SN host galaxies are in groups, and (ii) statistically, how much and how often do galaxy groups improve Hubble residuals.
From our pilot program with data from the AAT, we find 91\% of SN host galaxies to be in groups, with an estimated uncertainty of approximately 8\% from bootstrapping, while from the simulations, 73\% of SN host galaxies are found to be in groups.
SN host galaxies seem to be in groups more frequently than the complete distribution of galaxies, which may have a rate of $\sim$50--60\% in groups depending on the completeness of the galaxy sample analyzed.

According to our data, averaging the redshifts of galaxies in groups reduces the magnitude of the Hubble residual 50\% of the time.
That being said, galaxy groups result in large improvements in Hubble residuals much more frequently (15\% of the time) than worsening the Hubble residual by a large amount (5\% of the time).
From our simulations, we expect that the larger the group, the more likely grouping improves the Hubble residual.
With both the simulations and the data, we observe that the larger the Hubble residual is with the individual redshift, the more likely the group redshift will reduce the size of that Hubble residual.

In terms of scatter on the Hubble diagram for both the simulations and data, we observe the effect from using galaxy-group redshifts for both binned and unbinned samples.
The largest improvements in Hubble residual scatter come from the large groups at low redshift bins where we find the difference in STD values in quadrature, $\delta$, to be 0.120 mag in the simulations and 0.312 mag in the data.
For the unbinned sample from the data, $\delta=0.155$ mag which we find to be consistent with a bootstrapped sample from the simulations.
Using both simulations and data, we demonstrate that associating galaxies with galaxy groups reduces scatter on the Hubble diagram.

With the number of low-$z$ SNe Ia observed accelerating rapidly with recent samples (i.e., ATLAS, DEBASS, ZTF) and with LSST upcoming, plenty of galaxy groups for SN Ia hosts have yet to be defined.
With this analysis, we make the case for initiating a full-scale survey to define galaxy groups for all SN Ia hosts to improve measurements of cosmological redshifts.

\begin{acknowledgements}
\section*{Acknowledgements}
We thank the Templeton Foundation for directly supporting this research.
This work is based in part on data acquired at the Anglo-Australian Telescope under program A/2021A/16. We acknowledge the traditional custodians of the land on which the telescope stands, the Gamilaraay people, and pay our respects to elders past and present.

We thank Instituto de Astrofisica de Andalucia (IAA-CSIC), Centro de Supercomputacion de Galicia (CESGA), and the Spanish academic and research network (RedIRIS) in Spain for hosting Uchuu DR1, DR2, and DR3 in the Skies \& Universes site for cosmological simulations. The Uchuu simulations were carried out on Aterui II supercomputer at Center for Computational Astrophysics, CfCA, of National Astronomical Observatory of Japan, and the K computer at the RIKEN Advanced Institute for Computational Science. The Uchuu Data Releases efforts have made use of the skun@IAA\_RedIRIS and skun6@IAA computer facilities managed by the IAA-CSIC in Spain (MICINN EU-Feder grant EQC2018-004366-P).

D.S.~is supported by Department of Energy grant DE-SC0010007, the David and Lucile Packard Foundation, the Templeton Foundation, and Sloan Foundation. 
A.C.~and T.M.D.~acknowledge the support of an Australian Research Council (ARC) Australian Laureate Fellowship (FL180100168) funded by the Australian Government.
C.H.~acknowledges support from the ARC through project numbers DP2022010139 and CE230100016.
We would also like to thank Chris Lidman for their advice and input on this project as well as Samuel Hinton, Maddie Poole, Harry Van Der Ark, and Abbe Whitford for their help redshifting.
This research has made use of NASA’s Astrophysics Data System.

\section*{Software}
{astropy} \citep{astropy:2013,astropy:2018},
{matplotlib} \citep{Hunter07},
{Marz} \citep{Hinton2016Marz},
{numpy} \citep{numpy11}.
\end{acknowledgements}

\bibliographystyle{mn2e}
\bibliography{main}{}

\begin{appendix}
\section{Coordinate vs.~Velocity Figures for the Complete Sample from the AAT}\label{appendix:all_ra_dec_vs_velocity}

\begin{figure*}[!t]
    \centering
    \includegraphics[width=0.8\textwidth]{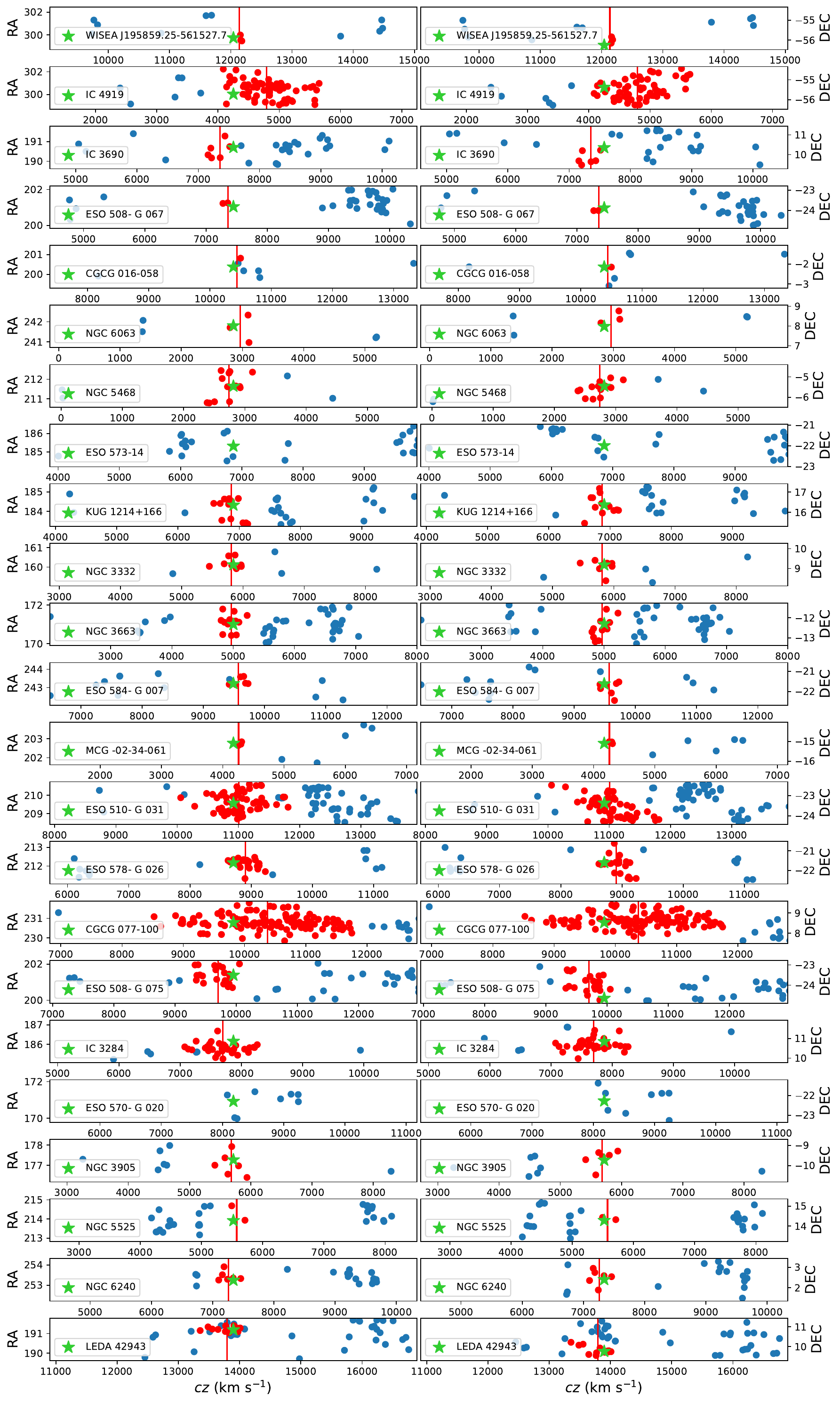}
    \caption{Similar to Fig.~\ref{fig:example_coord_vs_vel_groups}, coordinates and velocities from the AAT for all 23 SN host galaxies (green stars) and their nearby galaxies. Group members are in red and the group's average redshift is indicated by a red vertical line. 
    }
    \label{fig:coord_vs_vel_groups}
\end{figure*}

We provide galaxy-group depictions for the complete sample from the AAT in Fig.~\ref{fig:coord_vs_vel_groups}, which is the same as Fig.~\ref{fig:example_coord_vs_vel_groups}, but for all host galaxies in the sample.
Two host galaxies are not deemed to be in a galaxy group, ESO 573-14 and ESO 570- G 020.

\section{Cumulative Distributions for the Individual vs.~Group Hubble Residual}\label{appendix:cumul_dists}

\begin{figure}[!t]
    \centering
    \includegraphics[width=8.5cm]
    {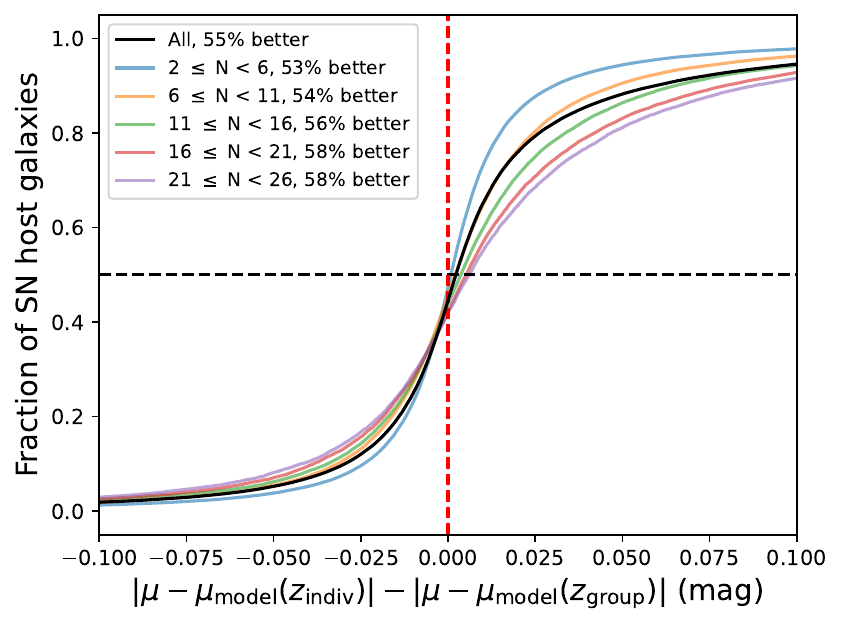}
    \includegraphics[width=8.5cm]{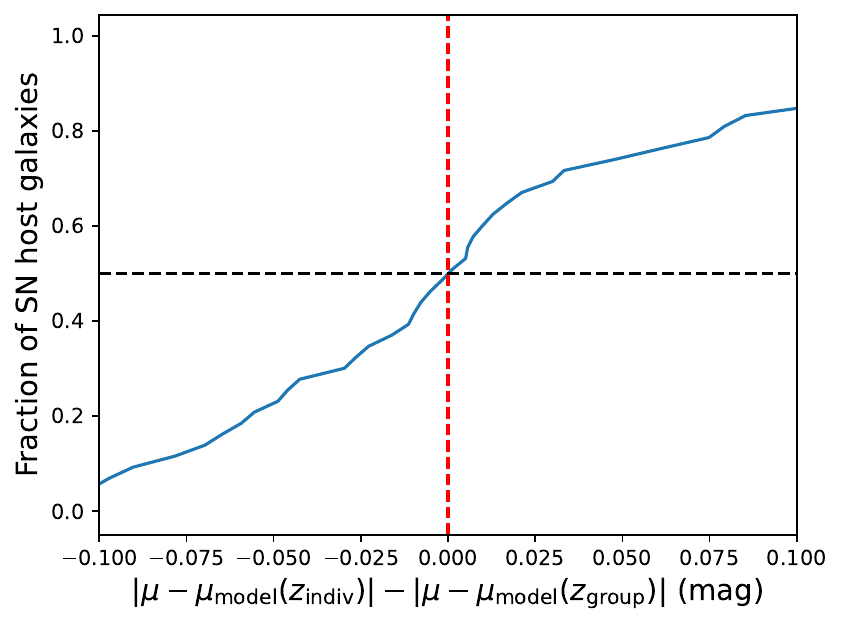}
    \caption{Simulated (left panel) and data (right panel) cumulative distributions of the difference in the magnitude of the Hubble residuals using the redshift of the individual galaxy vs.~the averaged redshift of the group. Bins for various group sizes are provided for the simulations in the left panel.}
    \label{fig:cumul_dist}
\end{figure}

With the simulations, we provide the expected cumulative distribution for the difference in the magnitude of the Hubble residual when using the individual redshift versus when using the group redshift in the left panel of Fig.~\ref{fig:cumul_dist}. 
A positive value on the x-axis indicates that the group redshift results in a smaller absolute Hubble residual compared to the individual redshift.
We see that more frequently, the group-averaged redshifts reduce the size of the Hubble residual. 
The proportion of Hubble residuals that are better with the group redshift increases with the number of group members, evolving from 53\% for groups with 2 to 5 members to 58\% for groups with 21 to 25 members. 
We attribute the frequency of reduced Hubble residuals from group redshifts being near 50\% to be due to the fact that other PVs, such as bulk motions, have not also been accounted for.
The simulations do not include intrinsic scatter in their Hubble residuals.

We provide the same figure but with data in the right panel of Fig.~\ref{fig:cumul_dist}.
For the data, the group redshift results in a smaller absolute Hubble residual than the individual redshift 50\% of the time.
That being said, groups result in large improvements in Hubble residuals ($>$0.1 mag) more often (15\% of the time) than making the residual larger/worse by a large amount (5\% of the time).
Comparing the cumulative distribution predicted by the simulations and that from the data is difficult since the group sizes are not perfectly analogous; the simulations tend to have slightly larger group sizes due to the inclusion of faint/dwarf galaxies.

\end{appendix}

\end{document}